\documentclass{article}
\begin{document}

\title{Space-time Torsion and Neutrino Oscillations in Vacuum}
\author{A. A. Sousa* , D. M. Oliveira and R. B. Pereira \\
%EndAName
Instituto de Ci\^{e}ncias Exatas e da Terra\\
Campus Universit\'{a}rio do Araguaia \\
Universidade Federal de Mato Grosso\\
78698-000 Pontal do Araguaia, MT, Brazil\\
}
\maketitle

\begin{abstract}
The objective of this study is to verify the consistency of the prescription
of alternative minimum coupling (connection) proposed by the Teleparallel
Equivalent to General Relativity (TEGR) for the Dirac equation. With this
aim, we studied the problem of neutrino oscillations in Weitzenb\"{o}ck
space-time in the Schwarzschild metric. In particular, we calculate the
phase dynamics of neutrinos. The relation of spin of the neutrino with the
space-time torsion is clarified through the determination of the phase
differences between spin eigenstates of the neutrinos.

PACS NUMBERS: 04.50.Kd, 04.20.Cv, 04.20.Fy

(*) E-mail: adellane@ufmt.br
\end{abstract}

\section{\textbf{Introduction}}

The description of the gravitational field in the Teleparallel Equivalent to
General Relativity (TEGR) introduced by Maluf led to tensorial expressions
for the energy, momentum and angular momentum of the gravitational field 
\cite{1}, \cite{2}. This theory can be considered as a reformulation of
Einstein's general relativity in terms of tetrad fields $e^{a}{}_{\mu },$
which is known as tetrad gravity.

We examine the consistency of the Dirac equation in the TEGR through a new
prescription of minimal coupling with the Dirac spinor fields $\psi .$ This
alternative prescription, involves the Levi-Civita connection $^{0}\omega
_{\mu ab}$ rather than spin connection of $\omega _{\mu ab}$ (field variable
independent of the tetrad field $e^{a}{}_{\mu })$. With respect to this
connection, Maluf showed in 1994 \cite{3} that it is possible to rule out $%
\omega _{\mu ab}$ both in Lagrangian and Hamiltonian formulation.

One motivation for this work, is the fact that, in the context of metric
affine theories of gravitation (MAG), Obkuhov and Pereira \cite{4} found an
inconsistency in the coupling of the Dirac spinor gravitational field when
the spin connection $\omega _{\mu ab}$ was used in the covariant derivative.
They found that the Dirac spinor fields couple with the gravitational field
in a manner consistent only with spin or matter and with the spin tensor
conserved. The TEGR was studied by Obkuhov and Pereira and can be considered
to be a special case of the general theory of MAG. In 2003, Maluf showed
that it is possible to bypass this problem with the use of a new type of
coupling to the Dirac spinor fields \cite{5}. To check the coupling
explicitly, we apply the Dirac equation of the TEGR with the connection $%
^{0}\omega _{\mu ab}$ (totally dependent on tetrad field) to the problem of
oscillations of solar neutrinos or mixing in vacuum. In particular, we
calculate the phase dynamics of neutrinos in space-time with torsion. The
contributions of torsion and their relationship with the directions of spin
in mass eigenstates were determined. These effects are compared with the
structure of Minkowski and Riemann-Cartan geometry. In the Riemann geometry,
there is no coupling between the spin of the particle and the gravitational
field.

The article is organized as follows: Section 2 presents a summary of the
Hamiltonian formulation of the TEGR and the problem of using the spin
connection $\omega _{\mu ab}$. In Section 3 we present the new minimum
coupling to the Dirac equation in the TEGR and apply this equation to model
neutrino oscillations in vacuum. The conclusions are presented in Section 4.

Notation: Space-time indices $\mu $,$\nu $, ... and local Lorentz SO(3, 1)
indices $a,b$,... run from 0 to 3. Time and space indices are indicated
according to $%
%TCIMACRO{\U{b5}}%
%BeginExpansion
{\mu}%
%EndExpansion
=0,$ $i$; $a=(0),$ $(i)$. The flat space-time is fixed by $\eta
_{ab}=e_{a\mu }e_{b\nu }g^{\mu \nu }=(-+++)$. The tetrad field $e^{a}{}_{\mu
}$ and the arbitrary spin affine connection $\omega _{\mu ab}$ yield the
usual definitions of the torsion and curvature tensors: $R_{ab\mu \upsilon
}=\partial _{\mu }\omega _{\nu ab}-\partial _{\nu }\omega _{\mu ab}-...,$ 
{\normalsize $T^{a}\,_{\mu \nu }=\partial _{\mu }e^{a}\,_{\nu }+\omega _{\mu
}{}^{a}\,_{b}$}$e^{b}\,_{\nu }-...$.The determinant of the tetrad field is
represented by $e=det(e^{a}{}_{\mu })$. $c$ is the speed of light, $\hbar $
is the Planck constant and $G$ is the Newtonian gravitational constant.

\section{\textbf{The Hamiltonian formulation of the TEGR}}

In 1994, Maluf \cite{3} implemented a Hamiltonian formulation of the TEGR
with local symmetry by imposing Schwinger%
%TCIMACRO{\U{b4}}%
%BeginExpansion
\'{}%
%EndExpansion
s \textit{time gauge} \cite{6} and found the following Hamiltonian density

\begin{eqnarray}
H(e_{\left( j\right) i},\Pi ^{\left( j\right) i},\omega
_{kab},P^{kab},\omega _{0ab}) &=&-N^{k}C_{k}-NC+\omega _{0ab}J^{ab}+
\label{20000} \\
&&-Ne\lambda ^{abik}R_{abik}-\Sigma _{(m)(n)}C^{\left( m\right) \left(
n\right) }+  \nonumber \\
&&+\partial _{k}\left( P^{kab}\omega _{0ab}\right) +\partial _{i}\left[
N_{k}\Pi ^{ki}+N(2eT^{i})\right] -  \nonumber \\
&&-\lambda ^{ij}\Pi _{\left[ ij\right] },  \nonumber
\end{eqnarray}%
where $\lambda ^{abik},$ $N^{k},$ $C^{\left( m\right) \left( n\right) }$ $%
J^{ab}$, $\lambda ^{ij}$and $N$ are Lagrange multipliers, $C_{k},$ $\Sigma
_{(m)(n)},$ $C,$ $\Pi _{\left[ ij\right] },$ $J^{ab},$ $R_{abik}$ are
constraints given in Ref.\cite{3}. $P^{kab}$ and $\Pi ^{\left( j\right) i}$
are components of the momentum canonically conjugate to $\omega _{kab}$ and $%
e_{\left( j\right) i}$, respectively.

Maluf showed that the constraints in Eq. $\left( \ref{20000}\right) $ are
first class by fixing $\omega _{0ab}=0.$ Then we can eliminate the momentum $%
P^{kab}$ from the Hamiltonian density. The evolution equation for $\omega
_{kab}$ leads to

\begin{equation}
\dot{\omega}_{kab}=\left\{ H,\omega _{kab}\right\} \mathbf{=}0,
\end{equation}%
and, for the equation to be consistent, it is necessary that $\omega
_{kab}=0 $ and then $\omega _{\mu ab}$ is ruled out of the theory. Then, the
Lagrangian density and the field equations are invariant under global
Lorentz transformations.

\section{Neutrino oscillations in the\ TEGR}

The tetrad field associated with the Schwarzschild metric was determined
using Schwinger's \textit{time gauge} \cite{6} $e^{\left( 0\right) }{}_{i}=0$%
, $e_{\left( k\right) }{}^{0}=0$ and the condition for spatial symmetry \cite%
{2}: $e_{(i)j}=e_{(j)i}$

\begin{equation}
e^{a}{}_{\mu }=\left( 
\begin{array}{cccc}
f & 0 & 0 & 0 \\ 
0 & f^{-1} & 0 & 0 \\ 
0 & 0 & r & 0 \\ 
0 & 0 & 0 & r\sin \theta%
\end{array}%
\right) ,  \label{eq39}
\end{equation}%
where $f=(1-\frac{2GM}{c^{2}r})^{\frac{1}{2}}$, $M$ is the mass of the sun.
The determinant of the tetrad field is given by $e=r^{2}\sin \theta .$

The equation proposed in the Maluf%
%TCIMACRO{\U{b4}}%
%BeginExpansion
\'{}%
%EndExpansion
s article \cite{5} for the minimum coupling is described by a covariant
derivative:%
\begin{equation}
D_{\mu }\psi =\partial _{\mu }\psi -\frac{i}{4}^{0}\omega _{\mu ab}\Sigma
^{ab}\psi .  \label{eq40}
\end{equation}

By substituting $(\ref{eq40})$ in the Dirac equation $i\hbar \gamma ^{\mu
}D_{\mu }\psi -m_{0}c\psi =0$ and using $^{0}\omega _{\mu ab}=-K_{\mu ab}$
where $K_{\mu ab}$ is the contortion tensor $K_{\mu ab}={\frac{1}{2}}%
e_{a}{}^{\lambda }e_{b}{}^{\nu }(T_{\lambda \mu \nu }+T_{\nu \lambda \mu
}-T_{\mu \nu \lambda }$), we have:

\begin{equation}
i\hbar \gamma ^{\mu }(\partial _{\mu }\psi +\frac{i}{4}K_{\mu ab}\Sigma
^{ab}\psi )-m_{0}c\psi =0.  \label{eq4040}
\end{equation}

By using the identity: $\epsilon _{abcf}\epsilon ^{abcd}=-3!\delta _{f}^{d},$%
and by considering only the axial part of the contortion tensor represented
by the vector $A^{a}=(A^{(0)},\vec{A})$, we obtain:%
\begin{equation}
K_{abc}\epsilon ^{abcd}=\epsilon _{abcf}A^{f}\epsilon ^{abcd}=\epsilon
_{abcf}\epsilon ^{abcd}A^{f}=-3!\delta _{f}^{d}A^{f}.  \label{5}
\end{equation}

By using the identities involving the Dirac matrices $\gamma
^{a}=e^{a}{}_{\mu }\gamma ^{\mu }=(\gamma ^{(0)},\gamma ^{(i)})$:

\begin{equation}
\gamma ^{a}[\gamma ^{b},\gamma ^{c}]=-2\eta ^{ab}\gamma ^{c}+2\eta
^{ac}\gamma ^{b}+2i\epsilon ^{abcd}\gamma _{\left( 5\right) }\gamma
^{(0)}\gamma ^{(1)}\gamma ^{(2)}\gamma ^{(3)}\gamma _{d},
\end{equation}

\begin{equation}
\gamma _{\left( 5\right) }=-i\gamma ^{(0)}\gamma ^{(1)}\gamma ^{(2)}\gamma
^{(3)},
\end{equation}

\begin{equation}
\Sigma ^{ab}=\frac{i}{2}\left[ \gamma ^{a},\gamma ^{b}\right] ,
\end{equation}

\begin{equation}
\gamma ^{(0)}\gamma _{(5)}\gamma _{a}A^{a}=\gamma _{(5)}A^{(0)}+%
\overrightarrow{\Sigma }\cdot \overrightarrow{A},
\end{equation}

\begin{equation}
\overrightarrow{\Sigma }=\left( 
\begin{array}{cc}
\overrightarrow{\sigma } & 0 \\ 
0 & \overrightarrow{\sigma }%
\end{array}%
\right) ,
\end{equation}

\begin{equation}
\gamma ^{a}\gamma ^{b}+\gamma ^{b}\gamma ^{a}=-2\eta ^{ab},
\end{equation}%
and by defining the components of the momentum as%
\begin{eqnarray}
p_{r} &=&-i\hbar \left[ \frac{\partial }{\partial r}-\frac{1}{r}-\frac{1}{2f}%
\frac{\partial }{\partial r}f\right] , \\
p_{\theta } &=&-\frac{i\hbar }{r}\left( \frac{\partial }{\partial \theta }-%
\frac{\cot \theta }{2}\right) , \\
p_{\varphi } &=&-\frac{i\hbar }{r\sin \theta }\frac{\partial }{\partial
\varphi },
\end{eqnarray}%
we can write the Dirac Hamiltonian matrix with the help of Eq. \ref{5}:

\begin{equation}
H=\left( 
\begin{array}{cc}
\begin{array}{c}
fIm_{0}c^{2}+\frac{3}{2}\hbar cf(\sigma ^{(1)}A^{(1)}+ \\ 
+\sigma ^{(2)}A^{(2)}+\sigma ^{(3)}A^{(3)})%
\end{array}
& 
\begin{array}{c}
f^{2}I\sigma ^{(1)}p_{r}c+fI\sigma ^{(2)}p_{\theta }c+ \\ 
+fI\sigma ^{(3)}p_{\varphi }c-\frac{3}{2}\hbar cfA^{(0)}I%
\end{array}
\\ 
\begin{array}{c}
f^{2}I\sigma ^{(1)}p_{r}c+fI\sigma ^{(2)}p_{\theta }c+ \\ 
+fI\sigma ^{(3)}p_{\varphi }c-\frac{3}{2}\hbar cfA^{(0)}I%
\end{array}
& 
\begin{array}{c}
-fIm_{0}c^{2}+\frac{3}{2}\hbar cf(\sigma ^{(1)}A^{(1)}+ \\ 
+\sigma ^{(2)}A^{(2)}+\sigma ^{(3)}A^{(3)})%
\end{array}%
\end{array}%
\right) ,  \label{eq43}
\end{equation}%
where $\sigma ^{\left( i\right) }$ are the Pauli matrices, and $I$ is the
identity matrix.

First, we analyze the azimuthal motion which, although is not realistic, is
more simple and was considered in Ref. \cite{7} and \cite{8} in a different
formalism.

\subsection{Azimuthal Motion}

Zhang and Pereira in Ref. \cite{9} was one of the first to calculate the
mass neutrino oscillation induced by torsion in the context of the weak
gravitational expansion method. For the calculation of the phase difference
in azimuthal and radial motion, we proceed in the same way as in Ref. \cite%
{10}. The Dirac Hamiltonian for azimuthal motion is given by $\left( \ref%
{eq43}\right) $, where we have substituted $\overrightarrow{p}%
=(p_{r},p_{\theta },p_{\varphi })=(0,0,p)$. Then the positive eigenvalue of
Eq. $\left( \ref{eq43}\right) $ is:

\[
E_{+}=p(1+\frac{m^{2}c^{2}}{2c^{2}p^{2}}-\frac{3\hbar \sigma ^{(3)}A^{(0)}}{%
2p})cf+\frac{3}{2}\hbar cf\vec{\sigma}\cdot \vec{A}, 
\]%
where we used the ultra-relativistic limit $E\simeq cp\gg mc^{2}.$ The
negative engenvalue gives similar results.

With the help of the unitary transformation for the spinor $\psi =U\left( 
\begin{array}{c}
\xi _{+} \\ 
\xi _{-}%
\end{array}%
\right) ,$ where $U$ is a the unitary matrix, we can write

\begin{equation}
E_{+}\xi _{+}=\left( fpc+\frac{fm^{2}c^{3}}{2p}\right) I\xi _{+}+
\label{eq46}
\end{equation}%
\[
+\left( 
\begin{array}{cc}
\frac{3}{2}\hbar cf\left( -A^{\left( 0\right) }+A^{\left( 3\right) }\right)
& \frac{3}{2}\hbar cf\left( A^{\left( 1\right) }-iA^{\left( 2\right) }\right)
\\ 
\frac{3}{2}\hbar cf\left( A^{\left( 1\right) }+iA^{\left( 2\right) }\right)
& -\frac{3}{2}\hbar cf\left( A^{\left( 0\right) }-A^{\left( 3\right) }\right)%
\end{array}%
\right) \xi _{+}=i\hbar \frac{\partial }{\partial t}\xi _{+}, 
\]%
where we found the following eigenvalues of Eq. $\left( \ref{eq46}\right) $
for the mass eigenstates with spin up and spin down

\begin{eqnarray}
\left\{ fpc+\frac{fm^{2}c^{3}}{2p}+\frac{3}{2}\hbar cfA\right\} \xi
_{+}^{\uparrow } &=&i\hbar \frac{\partial }{\partial t}\xi _{+}^{\uparrow },
\\
\left\{ fpc+\frac{fm^{2}c^{3}}{2p}-\frac{3}{2}\hbar cfA\right\} \xi
_{+}^{\downarrow } &=&i\hbar \frac{\partial }{\partial t}\xi
_{+}^{\downarrow },  \nonumber
\end{eqnarray}%
respectively. Here $A=\sqrt{(A^{(0)}-A^{(3)})^{2}+(A^{(1)})^{2}+(A^{(2)})^{2}%
}.$

The phase of the neutrino is obtained by the integration $\Phi =\frac{1}{%
\hbar }\int E_{+}dt,$ resulting in 
\begin{equation}
\Phi ^{\uparrow }=\frac{1}{\hbar }\left\{ \frac{fER}{c}\Delta \varphi +\frac{%
fm^{2}c^{3}R}{2E}\Delta \varphi +\frac{3}{2}\hbar fAR\Delta \varphi \right\}
.  \label{eq47.1}
\end{equation}

\begin{equation}
\Phi ^{\downarrow }=\frac{1}{\hbar }\left\{ \frac{fER}{c}\Delta \varphi +%
\frac{fm^{2}c^{3}R}{2E}\Delta \varphi -\frac{3}{2}\hbar fAR\Delta \varphi
\right\} .  \label{eq47.2}
\end{equation}%
for spin up and spin down, respectively. We used the ultra-relativistic
limit $pc\simeq E$ and $cdt\simeq Rd\varphi $, where $R$ is the radius for
the circular orbit of the neutrino and $\Delta \varphi $ is the angular
dislocation.

Now we have three possibilities

$1.$ From Eqs. $\left( \ref{eq47.1}\right) $ and $\left( \ref{eq47.2}\right) 
$, the phase difference between eigenstates that have the same spin state is
given by

\begin{equation}
\Delta \Phi =\Phi _{2}^{\uparrow }-\Phi _{1}^{\uparrow }=\frac{\Delta
m^{2}c^{3}}{2(E/f)\hbar }R\Delta \varphi .  \label{30000}
\end{equation}

$2.$ For the first mass eigenstate with spin down and the second eigenstate
with $spin$ up, we obtain

\begin{equation}
\Delta \Phi =\Phi _{2}^{\downarrow }-\Phi _{1}^{\uparrow }=\left\{ \frac{%
\Delta m^{2}c^{3}}{2(E/f)\hbar }-3fA\right\} R\Delta \varphi .  \label{eq53}
\end{equation}

$3.$ For the first mass eigenstate with spin up and the second eigenstate
with $spin$ down, we obtain

\begin{equation}
\Delta \Phi =\Phi _{2}^{\uparrow }-\Phi _{1}^{\downarrow }=\left\{ \frac{%
\Delta m^{2}c^{3}}{2(E/f)\hbar }+3fA\right\} R\Delta \varphi .  \label{eq54}
\end{equation}%
where $\Delta m^{2}=m_{2}^{2}-m_{1}^{2}.$

\subsection{The radial motion}

For the calculation of the phase difference in radial motion, we proceed in
the same way as in the case of azimuthal motion. Substituting $%
\overrightarrow{p}=(p_{r},p_{\theta },p_{\varphi })=(p,0,0)$ in the
expression $\left( \ref{eq43}\right) ,$ and using the ultra-relativistic
aproximation $pc\simeq E$, we obtain the following phase differences for the
neutrino mass eigenstates:

$1.$\ For eigenstates with the same spin:

\begin{equation}
\Delta \Phi =\Phi _{2}^{\uparrow }-\Phi _{1}^{\uparrow }=\frac{\Delta
m^{2}c^{3}}{2E\hbar }\Delta r.  \label{50000}
\end{equation}%
where $\Delta r=r_{B}-r_{A}.$ Here $r_{A}$ is the radius of the sun where
the neutrinos was produced and $r_{B}$ is the distance from the center of
the sun to the earth's surface where the neutrino was measured.

$2.$ For the first eigenstate with spin down and the second eigenstate with
spin up

\begin{equation}
\Delta \Phi =\Phi _{2}^{\downarrow }-\Phi _{1}^{\uparrow }=\frac{\Delta
m^{2}c^{3}}{2E\hbar }\Delta r-3A(\Delta r-\frac{MG}{c^{2}}\ln \frac{r_{B}}{%
r_{A}}).  \label{eq59}
\end{equation}

$3.$For the first eigenstate with spin up and the second eigenstate with
spin down

\begin{equation}
\Delta \Phi =\Phi _{2}^{\uparrow }-\Phi _{1}^{\downarrow }=\frac{\Delta
m^{2}c^{3}}{2E\hbar }\Delta r+3A(\Delta r-\frac{MG}{c^{2}}\ln \frac{r_{B}}{%
r_{A}}),  \label{eq60}
\end{equation}%
where $A=\sqrt{(A^{(0)}-A^{(1)})^{2}+(A^{(2)})^{2}+(A^{(3)})^{2}}.$ Here, we
used $cdt\simeq dr$, \ $f=(1-\frac{2MG}{rc^{2}})^{\frac{1}{2}}\simeq 1-\frac{%
MG}{rc^{2}}.$

Note that the magnitude of the contribution from the torsion is very small
compared with the mass differences between neutrinos \cite{10}.

If electron neutrinos are produced at time $t=0$, the time evolution of the
flavor eigenstate $\nu _{e}$ is given by:

\begin{equation}
\nu _{e}(t)=\cos \Theta e^{-i\Phi _{1}(t)}\nu _{1}+\sin \Theta e^{-i\Phi
_{2}(t)}\nu _{2},  \nonumber
\end{equation}%
where $\nu _{1}$ and $\nu _{2}$ ae called mass eigenstates at $t=0$ and $%
\Theta $ is called the \textquotedblleft mixture\textquotedblright\ angle.
The probability of measuring a muon or tau ineutrino is given by:

\begin{equation}
P(\nu _{e}\rightarrow \nu _{\mu ,\tau })=\sin ^{2}2\Theta \sin ^{2}\frac{%
\Delta \Phi }{2}.  \label{eq49}
\end{equation}

\section{\textbf{Conclusions}}

In azimuthal and radial motion, the torsion makes a contribution to the
phase dynamics of neutrinos that depends on the directions of spin in the
mass eigenstates. If both mass eigenstates have the same direction of spin,
there is no contribution to oscillation from torsion, but if the mass
eigenstates have opposite spin directions there is a contribution to the
oscillation of neutrinos that comes from torsion..

Without torsion or curvature (with $f=1$), our results coincide with results
for neutrino oscillations in vacuum in flat space-time \cite{7}.

Our result formally coincides with a result obtained by Adak $et$ $al.$ \cite%
{10}, from the Dirac equation in a Riemann-Cartan geometry with the
connection of spin $\omega _{\mu ab}=^{0}\omega _{\mu ab}+K_{\mu ab}.$
However, in radial motion, the radial component of momentum $p_{r}$ found in
our paper is less than the radial component of momentum found by Adak 
\textit{et al.} Therefore, we can claim that the energy of a neutrino in the
space-time of a teleparallel geometry, called Weitzenb\"{o}ck space-time, is
less than the energy of a neutrino in Riemann-Cartan space-time and the
probability of transition to a muon or tau neutrino is higher in the
teleparallel geometry.

The prescription of minimum coupling adopted by Maluf does not lead to
inconsistencies of the type that Obukhov and Pereira found and are
qualitatively consistent with the literature, at least for the problem of
neutrino oscillation. The formal inconsistency of the Dirac equation is
removed if we adopt the Levi-Civita connection of the theory as $^{0}\omega
_{\mu ab}=-K_{\mu ab}$ leaving the total Lagrangian density in the \textit{%
TEGR} in the presence of Dirac spinor fields invariant under global Lorentz
transformations.

\bigskip \noindent Acknowledgements

\noindent One of us (D. M. O.) would like to thank the Brazilian agency
CAPES for partial financial support.


\begin{thebibliography}{99}
\bibitem{1} J. W. Maluf, S. C. Ulhoa, F. F. Faria and J. F. da Rocha-Neto, 
\textit{Class. Quantum Grav}. \textbf{23}, 6245 (2006).

\bibitem{2} J. W. Maluf, J. F. da Rocha-Neto, T. M. L. Tor\'{\i}bio and K.
H. Castello-Branco, \textit{Phys. Rev. D }\textbf{65, }124001 (2002).

\bibitem{3} J. W. Maluf, \textit{J. Math. Phys., }\textbf{35}, 335, 1994.

\bibitem{4} Y. N. Obukhov and J. G. Pereira, \textit{Phys. Rev. D} \textbf{67%
}, 044016 (2003).

\bibitem{5} J. W. Maluf, \textit{Phys. Rev. D} \textbf{67}, 108501 (2003).

\bibitem{6} J. Schwinger, \textit{Phys. Rev}. \textbf{130}, 1253 (1963).

\bibitem{7} C. Y.Cardall and G. M. Fuller, \textit{Phys. Rev}\textbf{. D} 
\textbf{55}, 7960 (1997).

\bibitem{8} C. M. Zhang and A. Beesham, \textit{Gen. Rel. Grav}. \textbf{33,}
1011 (2001).

\bibitem{9} J. G. Pereira and C. M. Zhang, \textit{Gen. Rel. Grav}. \textbf{%
32,} 1633 (2000).

\bibitem{10} M. Adak, T. Dereli and L. H. Ryder, \textit{Class. Quantum Grav}%
. \textbf{18}, 1503 (2001).
\end{thebibliography}
\end{document}